\documentclass[a4paper,12pt]{article}  
\usepackage{latexsym}  
\usepackage{amsfonts} 
\usepackage{multirow} 
\usepackage{array} 
\usepackage{tabularx}  
\usepackage{amsmath}  
\usepackage[dvips]{graphicx}  
\newcommand{\be}{\begin{equation}}
\newcommand{\ee}{\end{equation}}
\newcommand{\bea}{\begin{eqnarray}}
\newcommand{\eea}{\end{eqnarray}}
\newcommand{\ba}{\begin{array}}
\newcommand{\ea}{\end{array}}

\def\bbox{{\,\lower0.9pt\vbox{\hrule \hbox{\vrule height 0.2 cm
\hskip 0.2 cm \vrule height 0.2 cm}\hrule}\,}}
\newcommand{\dsl}{\pa \kern-0.5em /}

\newcommand{\mysection}[1]{\section{#1}
            \setcounter{equation}{0}\setcounter{figure}{0}}

\def\5{\bar }  \def\6{\partial } \def\7{\tilde }
\def\8{\hat }

\newcommand{\beq}{\begin{equation}}
\newcommand{\eeq}{\end{equation}}

\newcommand{\bear}{\begin{eqnarray}}
\newcommand{\eear}{\end{eqnarray}}
\newcommand{\vx}{{\vec{x}}}

\newcommand{\vp}{{\vec{p}}}

\newcommand{\vk}{{\vec{k}}}
\newcommand{\vnabla}{{\vec{\nabla}}}

\begin{document}

\begin{titlepage}
\vfill
\begin{flushright}
UB-ECM-PF-05/00\\
CERN-TH-00-117\\
hep-th/0004115\\
\end{flushright}

\vfill

\begin{center}
\baselineskip=16pt
{\Large\bf Non-Relativistic Non-Commutative Field Theory and UV/IR
Mixing}
\vskip 0.3cm
{\large {\sl }}
\vskip 10.mm
{\bf ~Joaquim Gomis $^{\dagger,*,1}$,~Karl Landsteiner 
$^{\dagger,2}$, Esperanza Lopez $^{\dagger,3}$}\\
\vskip 1cm

{\small
$^\dagger$ Theory Division CERN\\
CH-1211 Geneva 23, Switzerland}\\
\vskip.5cm
{\small
$^*$
  Departament ECM, Facultat de F\'{\i}sica\\
  Universitat de Barcelona and Institut de F\'{\i}sica d'Altes
Energies,\\
  Diagonal 647, E-08028 Barcelona, Spain}\\ 
\end{center}
\vfill
\par
\begin{center}
{\bf ABSTRACT}
\end{center}
\begin{quote}
We study a non-commutative non-relativistic scalar field theory 
in $2+1$ dimensions. The theory shows the UV/IR mixing typical
of QFT on non-commutative spaces. The one-loop correction to
the two-point function turns out to be given by a $\delta$-function
in momentum space. The one-loop correction to the four-point function
is of logarithmic type. We also evaluate the thermodynamic potential
at two-loop order. To avoid an IR-singularity we have to introduce
a chemical potential. The suppression of the non-planar 
contribution with respect to the planar one turns out to depend
crucially on the value of the chemical potential.
\vfill
 \hrule width 5.cm
\vskip 2.mm
{\small
\noindent $^1$ E-mail: gomis@ecm.ub.es \\
\noindent $^2$ E-mail: Karl.Landsteiner@cern.ch\\
\noindent $^3$ E-mail: Esperanza.Lopez@cern.ch
}
\end{quote}
\end{titlepage}

\mysection{Introduction}
Non-commutative field theories have an
unconventional perturbative behaviour \cite{filk}-\cite{fischler2}. 
New infrared singularities in 
the correlation functions appear even for massive 
theories \cite{minwalla}-\cite{fischler2}. This phenomenon 
is due to an interplay  between the UV and IR induced by the 
Moyal phase appearing in the vertices.
Recently some  of the amplitudes of
these non-commutative theories have been derived
from string theory \cite{andreev}-\cite{joaquim2}.

In this note we analyze at the perturbative level the 
non-commutative version of a non-relativistic 
scalar field theory in $2+1$ dimensions
in order to gather more information about the UV/IR mixing
and the structure of degrees of freedom of non-commutative theories. 
Our motivation for investigating non-relativistic 
non-commutative field theories is twofold.
In non-relativistic quantum theory \linebreak non-commutativity of
space arises often in the effective description of
charged particles carrying dipole momentum moving in strong magnetic 
fields \cite{susskind}.
It seems natural then to look for the by now well-known
UV/IR mixing in the context of non-relativistic quantum field theory.
Another more theoretical motivation is that it might be easier
to understand the effects of non-commutativity of space in simpler
setups than relativistic quantum field theory. 
We also would like to emphasize that due to an ordering ambiguity
in the interaction vertices the particular model we are considering
can not be obtained as the non-relativistic limit of a relativistic
field model!

As we will see, also in this non-relativistic example
there is an interplay between the IR and UV behaviour due to the Moyal 
phases. For the two-point function a singularity of delta-function type 
appears. This should be contrasted with the pole like singularities
found for relativistic theories \cite{minwalla}. 
For the four-point function we find a singularity of a more familiar, 
logarithmic type. 

We study the system at finite temperature and non-zero chemical 
potential $\mu$. We compute the thermodynamical potential up to two loops. 
The presence of the chemical potential provides another scale besides the
non-commutativity scale and temperature. 
If $-\mu \!>>T$, the non-planar contribution is
strongly suppressed with respect to the planar one for thermal
wavelengths smaller than the non-commutativity scale.
This suggest a reduction of degrees of freedom running in the
non-planar graphs at high temperature \cite{fischler}\cite{fischler2}. 
The limit of $-\mu\!<<\! T$ is more involved. The non-planar graph
does not appear to be strongly suppressed as a function of the
temperature. It depends crucially on the ratio between the
chemical potential and the non-commutativity scale.

The paper is organized as follows. In section 2, we study the two and 
four-point function up to one-loop. In section 3 we  study the free 
energy up to two loops. We give some conclusions in section 4.

\mysection{Two and Four-Point Function at one-loop} 
\label{zeroT}

We will start by introducing the model. We will work in $2+1$ 
dimensions, where the non-commutativity affects only the spatial
directions. Non-commutative $R^2$ is defined by the commutation 
relations
\begin{equation}
\left[ x^\mu , x^\nu \right] = i \theta^{\mu\nu}\,,
\label{NC}
\end{equation}
with $\theta^{\mu\nu} = \theta \epsilon^{\mu\nu}$. The algebra of 
functions on non-commutative $R^2$ is defined through the star product
\begin{equation}
(f*g)(x) := \lim_{y\rightarrow x}
e^{\frac{i}{2}\theta^{\mu\nu} \partial^x_\mu\partial^y_\nu}f(x) g(y)
\label{starproduct}
\end{equation}
Here $x^\mu$ are taken to be ordinary c-numbers. 
We will study a self-interacting non-relativistic scalar field 
model, defined by the Lagrangian
\begin{equation}
{\cal L} = \phi^{\dagger} \left( i \partial_t + \frac{\vnabla^2}{2} \right)\phi - 
\frac{g}{4} \phi^{\dagger}*\phi^{\dagger}* \phi*\phi\,.
\label{model}
\end{equation}
The star product has been dropped in the term bilinear in the fields.
This is consistent since we can always delete one star in monomials
of fields in the action. This is equivalent to neglecting total
derivative terms.
In ordinary space-time this model arises as the low energy limit of a 
real relativistic scalar field with $\phi^4$ self interaction. It has been 
studied in \cite{bergman} as a model for applying renormalization 
to quantum mechanics with $\delta$-function potential \cite{jackiw}. 
The model is scale invariant in ordinary 
space-time since scale transformations in a non relativistic 
theory take the form $t \rightarrow \lambda^2 t, \vx \rightarrow \lambda 
\vx$. The scaling of $t$ is due to the fact that in (\ref{model}) the mass has
been scaled out by redefining $t \rightarrow m \: t$.  
It has been shown that the theory acquires a scale anomaly upon 
quantization quite analogous to what happens in relativistic quantum 
field theory. Of course, in the case considered here scale 
invariance is already broken at tree level by the non-commutativity scale $\sqrt{\theta}$. 

In going from ordinary space-time to the non-commutative one an ordering
ambiguity for the interaction term arises. We fix that ambiguity in 
(\ref{model}) by putting the $\phi^{\dagger}$ fields to the left.
The other possible ordering would have been to chose 
${\cal L}_{int}= -\frac{g'}{4}  \phi^{\dagger} * \phi * \phi^{\dagger} * \phi$.
A relativistic complex scalar field model with both interaction vertices 
has been considered in \cite{arefeva2}. There the authors showed that the 
theory 
was renormalizable at one-loop level only when $g=g'$ or 
$g=0$. We will later on show that no such restriction arises in the 
non-relativistic model (\ref{model}).

\begin{figure}[hb]  
\begin{center}  
\includegraphics[angle=0, width=0.4\textwidth]{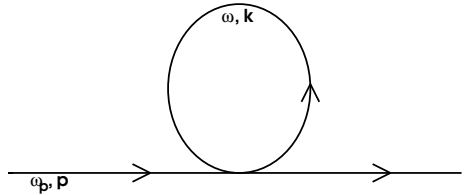}  
\caption{The tadpole contribution to the self-energy.}  
\label{tadpole}  
\end{center}  
\end{figure}

The solutions to the free field equations can be written as Fourier transforms
\begin{equation}
\begin{array}{lcl}
\phi(\vx,t)& = & \int \frac{d^2k}{(2\pi)^2}\, a(\vk)\, e^{-i(\omega_k t-\vk \vx)}\,, \\
\phi^\dagger(\vx,t) &=& \int \frac{d^2k}{(2\pi)^2}\, a^\dagger(\vk)\, e^{i(\omega_k t-\vk \vx)}\,,
\end{array}
\end{equation}
where $\omega_k = \frac{\vk^2}{2}$. 
The propagator of the theory is given by
\begin{equation}
\langle T \phi(x) \phi^\dagger (y) \rangle = 
\int \frac{d^2k d\omega}{(2\pi)^3} 
\frac{i \: e^{-i (\omega t - \vk \vx)}}{\omega-\vk^2 + i\epsilon} \,.
\end{equation}

We want to compute now the one-loop correction to the two-point function.
This is given by the tadpole diagram of figure \ref{tadpole}. In ordinary 
space-time we can employ a normal ordering prescription setting the 
tadpole to zero. In the non-commutative theory we expect a dependence of 
the tadpole on the external momentum due to the contribution of the 
non-planar diagrams. The planar and non-planar contributions are given by
\begin{equation}
-i \Sigma(E,\vp) = I_{planar} + I_{non-planar}\, ,
\end{equation}
and
\begin{equation}
\begin{array}{lcl}
I_{planar}& =& \frac{g}{2} \int \frac{d\omega d^2 k}{(2 \pi)^3}
\frac{1}{\omega - \frac{\vk^2}{2} +i\epsilon} \,\\[.5cm]
I_{non-planar}& =& \frac{g}{2} \int \frac{d\omega d^2 k}{(2\pi)^3}
\frac{\exp(i {\tilde p} k)}{\omega - \frac{\vk^2}{2} +i\epsilon}\,,
\end{array}
\end{equation}
where ${\tilde p}^{\mu}=\theta^{\mu \nu} p_{\nu}$.
In order to do the $\omega$-integration we recall that $(x+i\epsilon)^{-1} = 
{\cal P}\frac{1}{x} - i\pi \delta(x)$. This leaves us with the $\vk$-integrations
\begin{equation}
\begin{array}{lcl}
I_{planar} &=& \frac{-i g}{8\pi} \int_0^\Lambda k d k = \frac{-i g}{16\pi} \Lambda^2\, \\[.4cm]
I_{non-planar} &=& \frac{-i g}{16 \pi^2}\int d\varphi \int_0^\Lambda e^{i \tilde{p} k \cos(\varphi)} k d k = \frac{-i g}{8\pi \tilde{p}} \Lambda J_1(\tilde{p} \Lambda)\, ,
\end{array}
\end{equation}
where we introduced a UV-cutoff $\Lambda$ and $J_1(x)$ denotes a 
Bessel function. The quadratic divergence from the planar part
can be removed by adding a corresponding counterterm to the action,
${\cal L}_c= \delta \mu \phi^{\dagger} \phi$. 
The non-planar part reproduces the quadratic divergence for 
$\tilde{p}\rightarrow 0$ since $\lim_{x\rightarrow 0} \frac{J_1(x)}{x} = 
\frac{1}{2}$. In the limit $\Lambda\rightarrow \infty$, 
using $\int_0^\infty J_1(x) dx = 1$, 
it is straightforward to show that the result from the non-planar
diagram represents a delta-function in polar coordinates in 
$\tilde{p}$-space. We find then
\begin{equation}
\Sigma (E,p) = \frac{g}{4 \: \theta^2} \delta^2(\vp)\,.
\label{2point}
\end{equation}
Thus the situation is rather analogous to what happened in relativistic 
field theories. The limits of $\Lambda\rightarrow \infty$ and 
$p\rightarrow0$ do not commute. 

It is interesting to see that we can 
recover the delta-type singularity of the non-planar diagram as a limit
of the relativistic case. The relativistic theory is 
given by the Lagrange density
\begin{equation}
{\cal L}_{rel} = \frac{1}{2} (\partial\phi_{r})^2 - \frac{1}{2}m^2 \phi_r^2 - 
\frac{\lambda}{4!} \phi_r * \phi_r * \phi_r * \phi_r \,.
\label{relativistic}
\end{equation}
The non relativistic limit can be obtained as a $1/m$ expansion. We take 
off the fast oscillation due to the large mass and introduce dimensionless 
non relativistic fields by defining 
\begin{equation}
\phi_r = \frac{1}{\sqrt{2 m}} ( e^{-imt}\phi + e^{imt}\phi^\dagger)\,.
\label{relfield}
\end{equation}
To extract the non-relativistic limit we have to expand the vertex 
in \ref{relativistic} and compare with the vertex in \ref{model}. 
From the relativistic vertex we obtain
\begin{equation}
{\cal L}_{rel} = -\frac{\lambda}{4! m^2}\left(\phi^\dagger * 
\phi^\dagger * \phi * \phi + \frac{1}{2} \phi^\dagger * \phi * 
\phi^\dagger *  \phi\right)\,.
\label{vertex}
\end{equation}
Note that the non-relativistic limit produces both possible orderings in 
the interaction. Therefore in the non-commutative case our model \ref{model} is not the non-relativistic limit of a real relativistic scalar field. It 
turns out however that only the first vertex in \ref{vertex} contributes to 
the non-planar tadpole diagram. We should be able then to obtain the result
\ref{2point} from the relativistic case. Comparing \ref{model} with 
\ref{vertex} we see that $\lambda = 6 m g$ \footnote{Recall that in \ref{model} we
have rescaled $t \rightarrow m t$ and ${\cal L} \rightarrow m {\cal L}$ in
order to factor out the mass dependence.}. The non-planar contribution to 
the tadpole diagram in the relativistic theory is given by 
\begin{equation}
I_{rel} = \frac{\lambda}{6(2\pi)^3} \int d^3k \frac{e^{i\tilde{p}k}}{k^2-m^2} \,.
\end{equation}
In order to evaluate the integral we switch to Euclidean momentum and use Schwinger parameterization.
We obtain
\begin{equation}
I_{rel} = \frac{-i \lambda \pi^{\frac{3}{2}}}{6(2\pi)^3} \int_0^{\infty} \frac{d\alpha}{\alpha^{3/2}}
e^{-\frac{\tilde{p}^2}{4\alpha}-\alpha m^2} = \frac{-i \lambda e^{-\tilde{p} m}}{24\pi \tilde{p}}\,.
\end{equation}
There arises also a factor of $\frac{1}{2}$ which can be seen by noting that the integral $I_{rel}$
defines the relativistic self energy $\Sigma_{rel}$. The relativistic dispersion relation is
$(p_0-m)(p_0+m) - \vec{p}^2 - \Sigma_{rel}=0$. Setting $p_0 = m + E$ where $E$ is the 
non-relativistic 
energy we can go to the non relativistic limit by scaling $m\rightarrow \infty$ and
$E\rightarrow 0$ keeping the product $E m = \omega$ fixed. This is the non relativistic
energy of dimension $2$. In this way the relativistic dispersion relation becomes
twice the non relativistic one if we identify $\lim_{m\rightarrow \infty}\Sigma_{rel} = 2 \Sigma_{non-rel}$. Substituting for $\lambda$ it is then easy to
show that
\begin{equation}
\lim_{m\rightarrow \infty} I_{rel} = -i \frac{g}{2} \delta^2(\tilde{p})\,.
\end{equation}
Thus we reproduce precisely the non relativistic result (\ref{2point}).

If we formally sum all the tadpole diagrams contributing to the two-point 
function we obtain the following modified dispersion relation 
\begin{equation}
\omega = \frac{p^2}{2} + \frac{g}{4 \theta^2} \delta^{2}(\vec{p}).
\label{dispersion}
\end{equation}
In the resummation one encounters arbitrary high powers of $\delta$-functions.
Thus the resumation is highly ill defined. On the other hand we just showed that 
the dispersion relation \ref{dispersion} arises also as the limit of the relativistic one. 
Therefore we expect it to be correct on physical grounds. Alternatively we could keep the 
cutoff and arrive to \ref{dispersion} with a suitable smeared $\delta$-function. The meaning 
of \ref{dispersion} is that the energy of the zero momentum states is shifted by 
an infinite amount. However, it is important to note that the delta-function in
the dispersion relation is integrable. Thus wavepackets containing
zero momentum components still will have finite energy.

\begin{figure}[hb]   
\begin{center}  
\includegraphics[angle=0, width=0.7\textwidth]{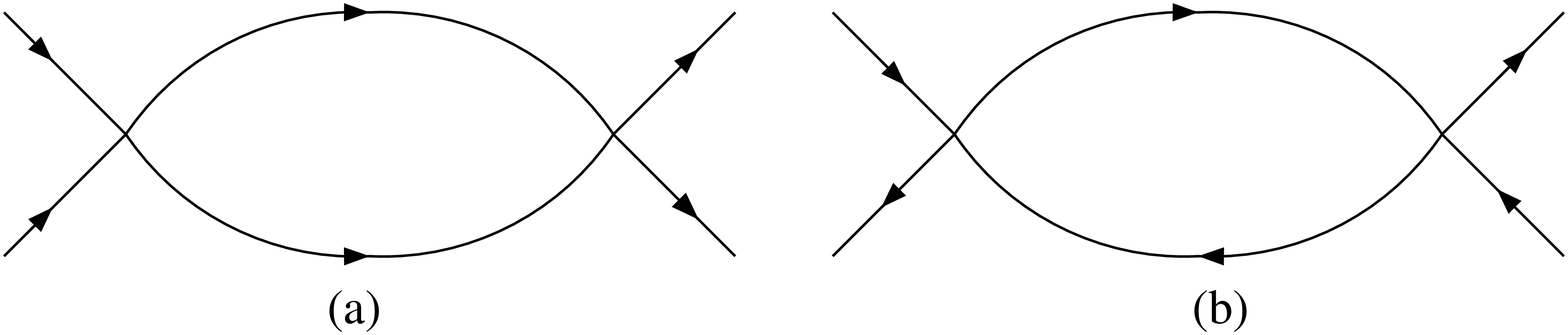}  
\caption{The one-loop contribution to the four-point function. The diagrams with momentum flow
as indicated in (b) vanish identically in non relativistic field theory.}  
\label{fourpoint}  
\end{center}  
\end{figure}

We would like now to evaluate the four-point function at one-loop.
In a non-relativistic theory only the {\it s} channel contributes, 
since the {\it t} and {\it u} channels contain internal lines flowing 
both forward and backwards in time and this evaluates to zero in a 
non-relativistic theory (see figure \ref{fourpoint}).  
As shown in \cite{arefeva2} the contributions from {\it u} and {\it t} channel
make the relativistic complex scalar field non-renormalizable if one does 
not also include the second possible ordering for the vertex. It is thus the
vanishing of u- and t- channel that allows us to ignore the second possible
ordering in the vertex. 
The non-relativistic one-loop 
four-point function is  \cite{bergman}
\begin{equation}  
\Gamma_4(\omega_i, \vec{p}_i,\Lambda)=
- \frac{\lambda^2}{8 \pi} \: \left( \mbox{log} \frac{\Lambda^2}{
E-\frac{P^2}{4}} + i \pi \right),
\label{4ord}    
\end{equation}  
where $E\!=\!\omega_1\!+\!\omega_2\!
=\!\omega'_1\!+\!\omega'_2$ and $P\!=\!p_1\!+\!p_2\!=\!q_1\!+\!q_2$
are the center of mass energy and momentum, and $\omega_i, p_i$ and 
$\omega'_i,p'_i$ the energy and momentum of the incoming and outcoming 
particles respectively; $\Lambda$ is an UV cutoff.

The one-loop four-point function for the non-commutative case is
given by 
\begin{equation}  
\Gamma_4 = 
 \frac{i \lambda^2}{2}  \cos \frac{{\tilde p}_1 p_2}{2}  
\cos \frac{{\tilde q}_1 q_2}{2}
\int \frac{d^2 k d \omega}{(2 \pi)^3} 
\frac{\mbox{cos}^2 \frac{{\tilde P} k}{2}}{(\omega - \frac{k^2}{2} + 
i \epsilon)(E -\omega - \frac{(k-P)^2}{2} + i \epsilon)},     
\end{equation}  
Using $\mbox{cos}^2 \: \frac{{\tilde P} k}{2} \!=\! \frac{1 \!+ 
\!\cos {\tilde P} k}{2}$ and writing $\cos {\tilde P} k$ in terms of 
exponentials, we can separate the planar and non-planar contributions. 
After doing the $\omega$ integration and shifting $k \rightarrow k + 
\frac{P}{2}$ we get for the non-planar part
\begin{equation}  
\Gamma_4^{non-planar}= - \frac{\lambda^2}{4}  \cos \frac{{\tilde p}_1 p_2}{2}  
\cos \frac{{\tilde q}_1 q_2}{2} \; \int \frac{d^2 k}{(2 \pi)^2} 
\frac{e^{i {\tilde P} k}}{k^2 - E+\frac{P^2}{4} - 2 i \epsilon}.  
\label{4noncom} 
\end{equation} 
This integral can be analyzed by changing to polar coordinates in 
momentum space. For angles such that ${\tilde P}. k \!>\!0$, we can
evaluate \ref{4noncom} by using a contour encircling the first
quadrant of the $|k|$-complex plane. For angles such that
${\tilde P}. k \!<\!0$, it is convenient to use a contour in the 
$|k|$-plane encircling the fourth quadrant. Adding both
contributions, we obtain the following result
\begin{equation}
\Gamma_4^{non-planar}=-\frac{\lambda^2}{16}  \cos \frac{{\tilde p}_1 p_2}{2}  
\cos \frac{{\tilde q}_1 q_2}{2} \left[
-Y_0\left({\tilde P} \sqrt{E-{\frac{P^2}{4}}}\right) +
i J_0 \left( {\tilde P} \sqrt{E - \frac{P^2}{4}}
\right)\right],
\label{fournc}
\end{equation}
where $J_0$ and $Y_0$ denote Bessel functions of first and second kind
respectively. In order to better
understand this expression, it is convenient to expand the Bessel
functions for small $\tilde P$. Up to a real constant and $O(\tilde{P})$
terms, the result is
\begin{equation}
\Gamma^{non-planar}_4=-\frac{\lambda^2}{16 \pi} \: \mbox{cos} 
\frac{{\tilde p}_1 p_2}{2} \: 
\mbox{cos} \frac{{\tilde q}_1 q_2}{2} \left(
\mbox{ln} \frac{ 1/ {\tilde P}^2}{E-\frac{P^2}{4}} +i \pi \right).
\end{equation}
The non-commutative phases regulate the otherwise divergent contribution
coming from high momentum. The resulting dependence of the non-planar 
diagram on the external momentum is smoother than for the two-point 
function. The external momentum $\tilde P$ acts as an UV cutoff
very much in the same way as in previously analyzed examples of 
relativistic theories \cite{minwalla}\cite{hayakawa}\cite{lenny}\cite{raamsdonk}.

\mysection{Finite Temperature Behaviour}

In this section we analyze the thermodynamics of our model. 
The physical reason to consider this system in a heat bath is to check
if there is  a reduction of degrees of freedom for the non-planar sector
of the theory. In the case of relativistic theories this was shown to happen
for thermal wavelengths smaller than the non-commutative length scale 
\cite{fischler}\cite{fischler2}  . 

Before we embark on doing the calculation we remind the reader of the 
following formula
\begin{equation}
\label{matsusum}
\sum_n \frac{1}{i \omega_n -x} = -\frac{\beta}{2}- 
\frac{\beta}{e^{\beta x}-1}\,.
\end{equation}
where $\omega_n = \frac{2\pi n}{\beta}$. The first term on the r.h.s. 
represents the zero temperature 
contributions. 
The resulting zero temperature
divergences can be canceled by the introduction of appropriate couterterms. 

We will compute the thermodynamic potential up to two loop. In order to cure
infrared divergences  we will introduce a chemical potential term 
$\mu \phi^\dagger\phi$ in our lagrangian \ref{model}. The introduction of
a chemical potential seems natural taking into account the renormalization
properties of the theory at zero temperature. Notice that now three scales
are present. Correspondingly we have two regimes of high temperature.
By high temperature we mean thermal wavelength \footnote{
The thermal wavelength of a non-relativistic system is given by $\lambda_T =
\frac{2 \pi}{\sqrt{T}}$ ($m=1$).} much smaller than the
scale of non commutativity, or equivalently, $T \theta >>1$.
The physics is then still dependent on the chemical potential\footnote{In non-relativistic
theory the chemical potential takes values in $(-\infty,0)$.}. In the regime
$-\mu \!>> \! T$ we expect a classical particle picture to be valid. We will also 
investigate the regime $T \! >> \! -\mu$, where classical field
theory is a good approximation to quantum statistical mechanics.

The one-loop contribution to the thermodynamic potential $F=-T \log Z$ is given by
\begin{equation}
 -T \int \frac{d^2k}{(2\pi)^2} \log\left(1-e^{-\beta(\frac{k^2}{2}-\mu)}\right) =
T^2 \mbox{Li}_2\left(1-e^{\frac{\mu}{T}}\right) \; ,
\label{oneloop}
\end{equation}
where $\mbox{Li}_2$ denotes the dilogarithm. The the two loop contribution is given by
\begin{equation}  
I=\frac{g}{2}\,T^2\,\sum_{l,n}\int \frac{d^2 p}{(2 \pi)^3} \frac{d^2 k}{(2 \pi)^3}
\frac{\cos^2 \frac{{\tilde p} \vk}{2}}{(i\omega_l - \frac{p^2}{2}+\mu)
(i\omega_n - \frac{k^2}{2}+\mu)}
\end{equation}  
As in the zero temperature case we substitute $\cos^2 \frac{{\tilde p} \vk}{2} = 
\frac{1+\cos {\tilde p} \vk}{2}$ separating the planar and non planar parts.

Using formula (\ref{matsusum}) we obtain three contributions to the 
planar part. The $(T=0,T=0)$ is a temperature independent divergence.
The $(T=0,T)$ contributions are divergent.
They can be canceled by adding counterterm of the form of the chemical 
potential $\delta\mu \: \phi^\dagger\phi$. The $(T,T)$ contribution can
be easily integrated  
\begin{equation}
I_{planar} = \frac{g}{8\pi^2} T^2 \left[\ln\left( 1-e^{\frac{\mu}{T}}\right)\right]^2\,.
\label{Tplanar}
\end{equation} 

The non-planar contribution to the free energy contains again three 
pieces. The first one is temperature independent and finite. The $(T=0,T)$ contribution is
\begin{equation}
I_{non-planar}^{T=0} = \frac{g}{2}  \int \frac{d^2 p}{(2\pi)^2}\frac{d^2k}{(2\pi)^2}
\frac{e^{i\tilde{p}k}}{e^{\beta(\frac{k^2}{2}-\mu)}-1} = 
\frac{g}{8\pi^2 \theta^2} \frac{1}{e^{-\frac{\mu}{T}}-1} \,.
\end{equation}
This can be interpreted as a one-loop contribution due to the shift in the
dispersion relation \ref{dispersion}. The $(T,T)$ contribution is
\begin{equation}
I_{non-planar}^{T} = \frac{g}{8\pi^2} \int p dp k dk \frac{J_0(\tilde{p} k)}
{(e^{\beta(\frac{k^2}{2}-\mu)}-1)(e^{\beta(\frac{k^2}{2}-\mu)}-1)}.
\label{2NPT}
\end{equation}

\noindent Since $J_0 \leq 1$, we see that the non-planar contribution is
suppressed with respect to the planar one. The strength
of the suppression will depend on the value of the two dimensionless
quantities $\theta T$ and $-\mu/T$.

We will analyze first the regime $-\mu /T >> 1$. In this limit
we can substitute the Bose-Einstein distribution by the
Maxwell-Boltzmann distribution. This corresponds to consider low
densities for the thermal gas. This is the particle 
approximation to the quantum field theory. 
In this limit we can evaluate the integral explicitly,
\begin{equation}
I_{non-planar}^T= \frac{g}{8\pi^2}\frac{T^2}{1+(\theta T)^2}e^{\frac{2\mu}{T}}\label{lowD}.
\end{equation}
For $\theta T \! <<\!1$ planar and non-planar graphs give the same
contribution. For $\theta T \!>>\! 1$ there is a very strong suppression
of the non-planar sector.
The $T^2$ dependence of \ref{Tplanar} is substituted by
$1/\theta^2$. When $T$ is larger than $1/\theta$ the thermal wavelength
$\lambda_T \sim 1/\sqrt{T}$ 
becomes smaller than the radius of a Moyal cell. 
Equation \ref{lowD} seems to indicate that the effective wavelength of 
the modes that circulate in the non-planar loop can not be smaller than 
the radius of the Moyal cell. 

We analyze now the
regime of small $-\frac{\mu}{T}\!<<\!1$. The classical thermal
field theory approximation consists in dimensionally
reducing the system along the Euclidean time direction, or equivalently,
considering only the zero mode in the sum over Matsubara frequencies.
In the limit of small $-\frac{\mu}{T}$ this approximation is valid up
to modes of momentum $k^2 \!< \! T$. On the other hand the 
non-commutative phases suppress modes of momentum $k^2 \! > \! \frac{1}{\theta}$
as can be explicitly seen from the Bessel function appearing in \ref{2NPT}. 
Therefore when $\theta T \! >> \! 1$ and $-\frac{\mu}{T}\!<<\!1$ we expect that the
classical field approximation will describe the leading behaviour
of the non-planar sector \cite{fischler2} \footnote{Notice that in our 
case the
two spatial directions are non-commutative. Therefore we expect
the suppression of high momenta by $\theta$ to be more effective
than in the cases studied in \cite{fischler2}, where the classical
approximation was applied to a system with odd spatial dimensions.}.
The integral \ref{2NPT} can be evaluated in this limit with the result
\begin{equation}
I^T_{non-planar}=\frac{g}{8 \pi^2} T^2 \mbox{G}\big( (-\mu \theta)^2 \big),
\end{equation}   
where $G(z)=G_{13}^{31}\big( z |_{000}^0 \big)
=
\frac{1}{2 \pi i} \int \Gamma(1+s)^3 \Gamma(-s) z^s ds$ denotes a Meijer 
G-function. The suppression of
the non-planar sector with respect to the planar one appears in this
case to be only logarithmic with the temperature. However, contrary to the 
previous case, the ratio between planar and non-planar contributions depends
also on $-\mu \theta$. For $-\mu \theta$ large the function $G$ tends to zero
implying an additional suppression for the non-planar sector. 
For $-\mu \theta$ small $G$ diverges. This divergence is associated 
to the infrared problems of the theory at small chemical potential. 

\mysection{Discussion and Conclusions}

We have seen that the phenomena of the UV/IR mixing is not only a 
characteristic of relativistic theories but also occurs in
non-relativistic theories. The model we have considered is a non-commutative
version of a 2+1 dimensional model that describes many particle quantum mechanics 
with a delta function interactions.
For the two-point function we have seen the appearance of IR singularity of
delta function type which changes the dispersion relation.
For the four-point function we found a logarithmic singularity.
Thus the non-relativistic model has an UV/IR mixing similar to the
relativistic field theories studied so far. Since our model
can not be embedded in a natural way in a string theory one might
interpret this as slight evidence that the IR singularities are not connected
to closed string states that do not decouple from the field theory.

The renormalizability of non commutative field theories to all loop order
is still an open problem \cite{grosse2}\cite{chepelev}. 
The non relativistic scalar field model 
might proof to be a simple and interesting toy model for such a study.
The fact that some diagrams vanish identically (such as t- and u-channel
contributions to the four-point function) could simplify a systematic
study of renormalizability. That in resumming the self-energy
insertions in the propagator one has to deal with powers of delta-functions
should not a priori be considered as a an unsurmountable obstacle.
As we argued such a formal resummation is physically well motivated.
Indeed the delta function appears also in the non relativistic limit
of the resummed propagator of relativistic $\phi^4$ theory. 

We have also studied the two loop correction to free energy and we have seen 
that the non-planar part of the theory is very sensitive to the value
of the chemical potential. At large negative values it turns out that
the non-planar part is strongly suppressed compared to the planar part.
In this regime the behaviour is similar to what has been found
in relativistic theories in \cite{fischler2}. The thermal wavelength
of the degrees of freedom in non-planar diagrams can not become smaller
that the non commutativity scale. Therefore these degrees of freedom
are suppressed at high temperature.

This interpretation is less clear at high temperature and small 
chemical potential. It turned out
that the non-planar part is at most logarithmically suppressed.
Given that these two regimes behave so differently it should be an interesting
direction of further research to study the effects of a chemical potential
also in relativistic, non-commutative field theories.

\section*{Acknowledgements}
We would like to thank Luis Alvarez-Gaum{\'e}, Herbert Balasin,
Jos{\'e} Barb{\'o}n, C{\'e}sar G{\'o}mez, Harald Grosse, Cristina Manuel, 
Antonio Pineda, Toni Rebhan and Miguel Angel Vazquez-Mozo for
helpful discussions. The work of J.G. is partially supported by 
AEN 98-0431, GC 1998SGR (CIRIT). K.L. and E.L. would like to thank
the Erwin Schr{\"o}dinger Institut for 
Mathematical Physics, Vienna for 
its hospitality.

\end{document}